\documentclass[pra,twocolumn,showpacs,floatfix,showkeys]{revtex4-1}
\usepackage{graphicx}
\usepackage{dcolumn}
\usepackage{bm}
\usepackage{amsmath}
\usepackage{amssymb}
\usepackage{revsymb4-1}

\begin{document}

\title{Gap solitons under competing local and nonlocal nonlinearities}
\author{Kuan-Hsien Kuo$^1$, YuanYao Lin$^1$, Ray-Kuang Lee$^1$, and
Boris A. Malomed$^2$} \affiliation{$^1$Institute of Photonics
Technologies, National Tsing-Hua University,
Hsinchu 300, Taiwan\\
$^2$Department of Physical Electronics, School of Electrical
Engineering, Faculty of Engineering, Tel Aviv University, Tel Aviv
69978, Israel.}

\begin{abstract}
We analyze the existence, bifurcations, and shape transformations of
one-dimensional gap solitons (GSs) in the first finite bandgap
induced by a periodic potential built into materials with local
self-focusing and nonlocal self-defocusing nonlinearities.
Originally stable on-site GS modes become unstable near the upper
edge of the bandgap with the introduction of the nonlocal
self-defocusing nonlinearity with a small nonlocality radius .
Unstable off-site GSs bifurcate into a new branch featuring
single-humped, double-humped, and flat-top modes due to the
competition between local and nonlocal nonlinearities. The mechanism
underlying the complex bifurcation pattern and cutoff effects
(termination of some bifurcation branches) is illustrated in terms
of the shape transformation under the action of the varying degree
of the nonlocality. The results of this work suggest a possibility
of optical-signal processing by means of the competing nonlocal and
local nonlinearities.
\end{abstract}

\maketitle



\section{INTRODUCTION}

\label{sec1} The concept of photonic crystals (PhCs), i.e., artificial media
with a periodic modulation of local optical characteristics, provides ways
to tailor the dispersion, diffraction, and routing of electromagnetic waves~%
\cite{book1}. As for natural crystals, fundamental characteristics of the
PhCs are the band-diagrams, which reveal gaps where Bloch waves cannot
propagate. In PhCs made of nonlinear materials, self-trapped localized modes
in the form of the gap solitons (GSs) may exist, as a result of the
interplay of the Kerr-type nonlinearity and periodic structures~\cite%
{npc-book,Krug,Ostrovskaya03, Konotop}. Unlike spatial bright solitons
supported by the balance between the self-focusing nonlinearity and
diffraction in uniform bulk media~\cite{book2}, the dispersion relation
induced by the PhC makes it possible to create GSs in both focusing and
defocusing media. Combining assets of PhCs and regular solitons, GSs have a
potential for applications to soliton-driven photonics. New technologies
enabling reconfigurable optical lattices, such as photorefractive crystals
\cite{Efremidis2002} and nematic liquid crystals \cite{Peccianti2002}, also
open new ways to control the dynamics of solitary waves by adjusting the
lattice depth and period.

While the modulational ~\cite{bifur-gap-2004} and oscillatory instabilities
\cite{osc-ins-1994,osc-ins-1998} impose limits on the use of the GSs, the
stability and mobility of the GSs may be enhanced in nonlinear media
featuring a nonlocal response~\cite{xu2005,joa-yylin,pra-yylin2,
pra-10-yylin}. The nonlocal nonlinearity is important when the correlation
radius of the material's response function becomes comparable to the
transverse width of the wave packet~\cite{access-soliton, Wieslaw2000}. The
nonlocal nonlinearity gives rise to specific features in the soliton
dynamics, including the modification of the modulational~\cite{Krolikowski04}%
, azimuthal~\cite{Anton06}, and transverse \cite{TI-YY} instabilities.
Suppression of the collapse of multidimensional solitons \cite{Bang02}, a
change of the character of interactions between them \cite{Peccianti02},
formation of soliton bound states \cite{Torner05}, merger of colliding
solitons into a standing wave \cite{nbragg-YY}, and families of dark-bright
soliton pairs \cite{nlocal-YY} were predicted recently too. Experimental
observations of the nonlocal response have also been demonstrated in sundry
media, including photorefractive media~\cite{Duree93}, nematic liquid
crystals \cite{Conti03}, and thermo-optical materials \cite%
{Rotschild05,Efremidis2008}, with a large tunable range of the nonlocality
degree.

In this work, we aim to study GS modes in the first bandgap of the
model including local self-focusing and nonlocal self-defocusing
nonlinearities. When the nonlocality radius is zero, we assume equal
magnitudes of the self-focusing and self-defocusing terms, i.e.,
complete cancelation of the nonlinearity, hence no solutions in the
bandgap. For a small degree of the nonlocality, the existence,
bifurcation, and shape transitions of the emerging bright GSs are
analyzed. With the competing local and nonlocal interactions of
opposite signs, the family of on-site GS modes remain stable,
obeying the "anti-Vakhitov-Kolokolov" criterion \cite%
{antiVK}, for the case that the nonlocal perturbation of the refractive
index is small, but become unstable near the upper edge of the bandgap.
However, the bifurcation generating the GSs near the other edge of
the bandgap features an inverted slope of the bifurcation curve for a
relatively small degree of the nonlocality, in which case the on-site GSs
are unstable. The off-site GS family features single-humped, double-humped,
and flat-top profiles for different degrees of the nonlocality. We also
investigate the situation in the space of the soliton's propagation constant
and power, for the varying nonlocality degree, in order to illustrate the
mechanism underlying the complex bifurcation pattern and related cutoff
effects (termination of some solution branches). Using results reported in
this work, we discuss the possibility to design GS-based signal-processing
schemes by dint of manipulating the nonlocal interactions.

In addition to optics, GSs of matter waves have also been theoretically
studied~\cite{gap-wave} and experimentally created~\cite{gap-BEC-exp} in
Bose-Einstein condensates (BECs) formed by atoms with repulsive
interactions, trapped in optical-lattice (OL) potentials. In addition to the
known contact interaction in the BECs of alkali-metal atoms, the interaction
of chromium atoms, $^{52}$Cr, includes a dipole-dipole interaction, which is
intrinsically anisotropic and nonlocal. The condensate of $^{52}$Cr was
created and investigated using magnetic \cite{Pfau, Werner, Stuhler,
Griesmaier} and all-optical \cite{optical} traps, see also review \cite{dip}%
. By adjusting the orientation of the dipoles, one can effectively control
the nonlocal dipole-dipole interactions. For the dipolar BEC trapped in OLs,
the competition between the contact and long-ranged dipole-dipole
interactions not only dramatically change the band structures of nonlinear
Bloch waves~\cite{pra78-yylin}, but also modifies families of matter-wave
solitons~\cite{cuevas2009}.

The rest of the paper is organized as follows. In Sec.~\ref{sec2}, the model
including the competing local and nonlocal nonlinearities is described.
Properties of the on-site and off-site GS families supported by the local
self-focusing nonlinearity are recapitulated to show a transition of mode
profiles in the first bandgap. Results produced by the interplay of the
local self-focusing and nonlocal self-defocusing nonlinearities for on-site
and off-site GS families are reported in Sections~\ref{sec4} and \ref{sec5},
respectively. Tracing the change of the corresponding GS shapes in the
parameter planes, we explain the character of the corresponding bifurcation
and identify a possible control mechanism for the optical-signal processing.
Section~\ref{sec6} concludes this work.

\section{The gap-soliton family with the local self-focusing nonlinearity}

\label{sec2} Considering a wave packet propagating along the $\eta $ axis in
the nonlinear PhC structure, we assume that the embedded medium gives rise
to two kinds of the nonlinearity simultaneously. This system is modeled by
the modified nonlinear Schr\"{o}dinger equation~\cite{pra-10-yylin},
\begin{eqnarray}
&&i\frac{\partial \Psi }{\partial \eta }=-\frac{1}{2}\frac{\partial ^{2}}{%
\partial \xi ^{2}}\Psi +V(\xi )\Psi +\sigma n(\xi )\Psi +\rho |\Psi
|^{2}\Psi ,  \label{eq:gp1} \\
&&n-d\frac{\partial ^{2}n}{\partial \xi ^{2}}=|\Psi |^{2},  \label{eq:n}
\end{eqnarray}%
where $\Psi (\eta ,\xi )$ is the slowly varying amplitude of the electric
field, $V(\xi )$ is the periodic potential, and a perturbation of the
refractive index, $n(\xi )$, accounts for the diffusive nonlinear response
with the nonlocality degree (which scales as a squared nonlocality radius)
designated by parameter $d$. Sign parameters $\sigma ,\rho =+1$ and $-1$
correspond to the self-defocusing and self-focusing nonlinearities,
respectively. Below, we fix $\sigma =+1$ and $\rho =-1$ for a system with
the nonlocal self-defocusing and local self-focusing nonlinearities.

\begin{figure}[tbp]
\centering
\includegraphics[width=8.3cm]{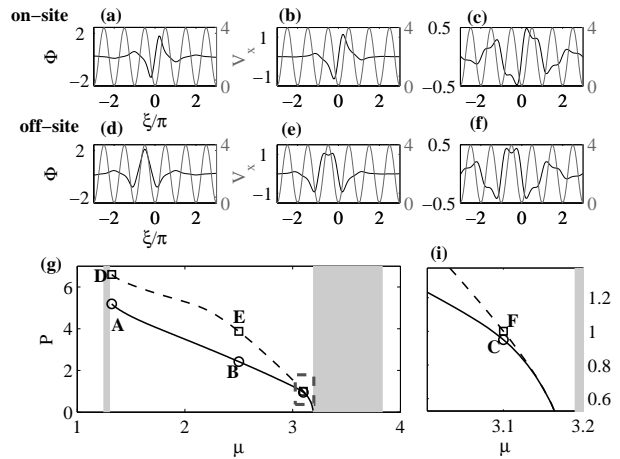}
\caption{Examples of on-site (a-c) and off-site (d-f) gap-soliton solutions
and the corresponding potential $V$ in the case of the local self-focusing
Kerr nonlinearity, depicted by black and gray lines, respectively. The
propagation constant of each mode is indicated by labels A through F in
panel (g), where the relations between the propagation constant and power
are plotted by the solid and dashed curves, respectively, for the on-site
and off-site modes. Panel (i) shows the blow-up of the area enclosed by the
dashed box in (g). Here and in other figures, shaded areas represent Bloch
bands bordering the first finite bandgap.}
\label{fig:mode-d0}
\end{figure}

Stationary solutions with propagation constant $-\mu $ are looked for as $%
\Psi (\eta ,\xi )=\Phi (\xi )e^{-i\mu \eta }$. For periodic potential $V(\xi
)=4\mathrm{sin}^{2}(k_{0}x)$ with $k_{0}=1$ fixed by rescaling, the
linearized version of Eq. (\ref{eq:gp1}) gives rise to the bandgap
structure, with the first finite bandgap being $1.3047<\mu <3.1896$. In this
bandgap, GSs can exist in form of on-site and off-site modes being supported
solely by the local nonlinearity, i.e., with $\sigma =0$~\cite%
{bifur-gap-2004}. For the self-focusing Kerr nonlinearity, we display
examples of both the on-site and off-site GS solutions in Fig.~\ref%
{fig:mode-d0}. The relationships between $\mu $ and the power of these
modes, $P=\int_{\infty }^{\infty }{d\xi }|\Phi |^{2}$, are shown in the
form of the bifurcation curves in Fig.~\ref{fig:mode-d0}(g), with labels A
through F referring to typical GS mode profiles in subplots (a-f).

Under the self-focusing nonlinearity, these GSs bifurcate from the upper
edge of the first bandgap (in other words, from the lower edge of the second
finite Bloch band), where the spatial-dispersion law features the anomalous
sign~\cite{perl2003}. In the entire first bandgap, the odd on-site GS modes,
shown in Fig.~\ref{fig:mode-d0}(a-c), feature two major peaks within one
lattice cell . Tails of these \ modes become conspicuous when the
propagation constant moves close to the upper edge of the bandgap. On the
other hand, a shape transition is demonstrated by the off-site even GS
modes. Near the lower edge of the bandgap, the GS solution has a single
major peak coinciding with a local maximum of the periodic potential, as
shown in Fig.~\ref{fig:mode-d0}(d). By tracing the variation of the
propagation constant and power along the relation for these off-site modes
shown by the dashed line of Fig.~\ref{fig:mode-d0}(g) to point F, it is seen
in Fig.~\ref{fig:mode-d0}(f) that the center of the modal profile breaks
into two peaks. The double-peaked solution is formed due to the balance
between the repulsive potential barrier and self-trapping induced by the
Kerr nonlinearity, as the characteristic self-trapping length becomes larger
than a half of the lattice period (roughly equivalent to the width of the
barrier potential). In between, we observe a smooth shape transition of the
GS from the single-humped shape in Fig.~\ref{fig:mode-d0}(d) to a nearly
flat-top one in Fig.~\ref{fig:mode-d0}(e), and, finally to the double-peaked
mode in Fig.~\ref{fig:mode-d0}(f).

The off-site GSs exist when the nonlinear self-trapping is stronger than the
repulsion induced by the potential barrier, giving rise to an effective a
potential well holding the localized modes, as seen in Figs.~\ref%
{fig:mode-d0}(d-f). On the other hand, the effect of the lattice potential
is stronger than that of the nonlinearity in the case of the on-site modes,
which, together with the contribution of the gradient energy, determines
their shapes in  Fig.~\ref{fig:mode-d0}(a-c). When the refractive index
correction gets weaker in accordance with the power reduction, the concave
lattice potential create a potential barrier to tailor and split the wave
function into a form of the states that similar to a binding profile from
two on-site modes in Fig.~\ref{fig:mode-d0}(b).

\section{On-site gap solitons under competing local and nonlocal
nonlinearities}

\label{sec4} 
\begin{figure}[tbp]
\centering
\includegraphics[width=8.3cm]{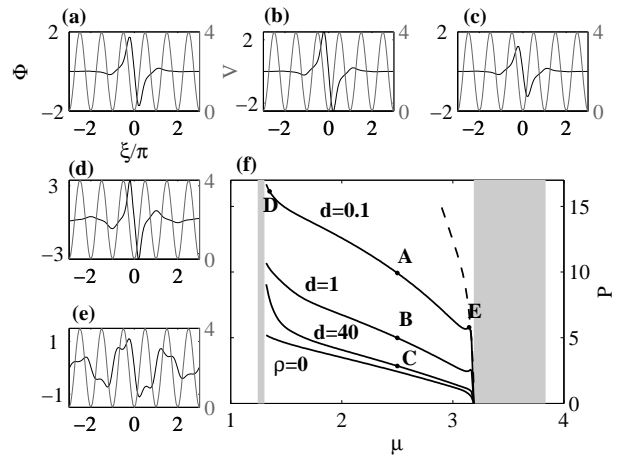}
\caption{Examples of on-site (a-e) gap solitons in the case of the competing
local and nonlocal nonlinearities (black lines) and the corresponding
potential $V$ (gray lines), for different degrees of the non-locality $%
d=0.1,1,40$, and the case of $\protect\rho =0$ (local only). The
corresponding values of propagation constant $\protect\mu $ are
labeled
(A-E) in panel (f). The dashed line in (f) is the asymptotic curve for $%
d=0.1 $. }
\label{fig:cn-onsite}
\end{figure}

In this section, we introduce the self-defocusing nonlocal
nonlinearity, setting $\sigma =1$ in Eq. (\ref{eq:gp1}). Obviously, in this
case the total nonlinearity cancels out to zero in the limit of $d=0$. At $%
d>0$, the overall nonlinearity is a self-focusing, because the diffusive
nonlocal kernel produces a spatially wider and less intensive perturbation
of the nonlinear refractive index, in comparison with that corresponding to
the local nonlinearity. Then, similar to the situation in the linear model
outlined above, one may expect the corresponding GS modes to bifurcate from
the upper edge of the first bandgap, where the effective spatial dispersion
is anomalous.\ In Fig.~\ref{fig:cn-onsite}, we show examples of on-site GSs
supported by the competing local focusing and nonlocal defocusing
nonlinearities for different degrees of non-locality, $d=0.1,1,40$, and the
case of $\rho =0$ (the local nonlinearity only)for the comparison.

For different degrees of the nonlocality, the profiles of the on-site GS
solutions vary slightly, remaining similar to their counterparts in the
local model. On the other hand, Fig.~\ref{fig:cn-onsite}(f) demonstrates
that the power required for the formation of the GSs is higher in the model
with the competing nonlinearities than in the local one because the net
nonlinearity is effectively reduced by the competition of the nonlocal
nonlinear response with the local Kerr term. In the limit of the ultimate
nonlocality, $d\rightarrow \infty $, the $P(\mu)$ curve converges to that
in the local model (with $\rho =0$). The latter feature is explained by the
fact that, in this limit (which corresponds to the model of the so-called
"accessible solitons" \cite{access-soliton}), the nonlocal
nonlinear response amounts to a weak constant background, which shifts the
propagation constant by the amount proportional to $P/\sqrt{d}$~\cite%
{pra-10-yylin}.

\begin{figure}[tbp]
\centering
\includegraphics[width=8cm]{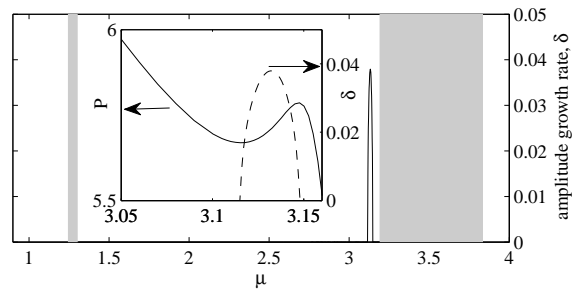}
\caption{Amplitude growth rate of small-perturbation modes upon the gap solitons in the case of the competing local and nonlocal nonlinearities for the degrees of the non-locality $%
d=0.1$. The inset shows the blow-up of stability analysis along with the corresponding power-propagation constant relations.}
\label{fig:cn-onsite-stab}
\end{figure}

In Fig.~\ref{fig:cn-onsite}(f), the bifurcation curve shows an
abrupt change of the slope both for $d=1$ and $d=0.1$ (in the latter
case, near point E). In a small region to the left of this point,
the slope changes its sign to $\text{d}\mu /\text{d}P>0$, and the corresponding on-site GSs are unstable. 
It is known that the bifurcation curves for GS families supported by
the local self-focusing nonlinearity obeys an inverted
"anti-Vakhitov-Kolokolov" criterion \cite{bifur-gap-2004},
$\text{d}\mu /\text{d}P<0$, under which the on-site GSs are stable
(see also Ref. \cite{antiVK}). To elucidate, we analyzed the stability of the
numerically found GS families by considering small-perturbation
modes and calculating their eigenvalues. %
In specific, we show in Fig.~\ref{fig:cn-onsite-stab} the linear stability analysis spectrum over the entire first band gap regime. It is clearly illustrated in the inset of Fig.~\ref{fig:cn-onsite-stab}, unstable small-perturbation eigenmodes are obtained only within the region where the power dependence $\text{d}\mu /\text{d}P>0$ and therefore confirms the inverted "anti-Vakhitov-Kolokolov" criterion. The evolution of these on-site gap solitons generated by direct numerical simulations further shows the collapse of on-site solitons falling out of the  "anti-Vakhitov-Kolokolov" regime. 

When the propagation constant of the GS is close to the upper edge of the bandgap
(on the right-hand side of the slope-change point), the slope of the $P(\mu )
$ becomes negative again. In that case, the GS is broad, spanning a few
lattice periods, and resembles \emph{gap wave} modes~\cite{gap-wave}, as
seen in Fig.~\ref{fig:cn-onsite}(e). The abrupt slope change smooths out
with the increase of the nonlocality degree, disappearing at $d\simeq 5$,
due to the fact that the nonlocal perturbation of the refractive index
becomes small for the strong nonlocality.

For a sufficiently small degree of the nonlocality, $d$, Fig.~\ref%
{fig:cn-onsite} shows that the power required to form the GSs increases with
the decrease of $d$, diverging at $d\rightarrow 0$. In the regime of the
weak nonlocality, the effective competing nonlinearity can be approximated,
to the first order in $d$, as~\cite{nlocal-YY} $n(\xi )|\Psi |^{2}-|\Psi
|^{2}\Psi \approx d\left( |\Psi |^{2}\right) _{\xi \xi }$, which implies
that the soliton's power scales as $1/d$. An asymptotic curve based on this
approximation is shown by the dashed line in Fig.~\ref{fig:cn-onsite}(f), to
illustrate the bifurcation of the on-site GSs near the upper edge of the
first bandgap.

\section{Off-site gap solitons under competing local and nonlocal
nonlinearities}

\label{sec5} 
\begin{figure}[tbp]
\centering\includegraphics[width=8.3cm]{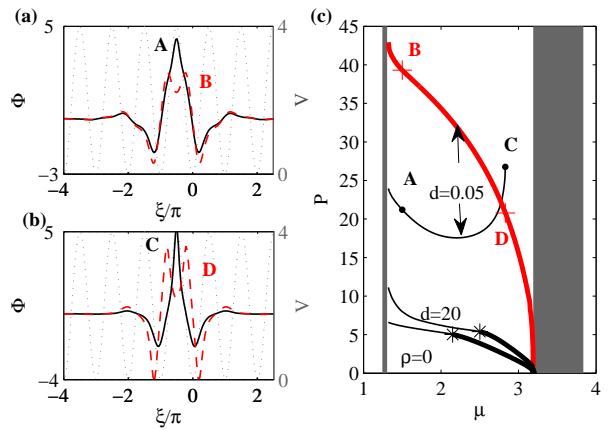}
\caption{(Color online) Off-site gap solitons in the model with the
competing local self-focusing and nonlocal self-defocusing nonlinearities
for the points marked in panel (c): A and B (a), C and D (b). The
bifurcation curves for two very different values of the degree of the
nonlocality, $d=0.05$ and $20$, are shown in (c) to indicate the cases with
and without the new branch.Thin and thick portions of $P(\protect\mu )$
curves pertain to the solitons with single- and double-peak profiles,
respectively.}
\label{fig:cn-offsite}
\end{figure}

Next, we aim to study off-site GSs in the first finite bandgap
under the action of the competing nonlinearities. As mentioned in Section~%
\ref{sec2}, the profile of the off-site GS solutions changes from
single-humped to double-humped as propagation constant $\mu$ approaches the
upper edge of the bandgap. To indicate the change of the profile caused by
the introduction of the competing nonlocal nonlinearity, in Fig.~\ref%
{fig:cn-offsite}(c) we use thin and thick lines to distinguish portions of
the $P(\mu )$ curves representing such single- and double-humped profiles.
Again, in the limit of the strong nonlocality, the correction to the
refractive index induced by the nonlocal nonlinearity is widely spread in
the space and very small with respect to the effect of the Kerr
nonlinearity, which makes the competition negligible. For example, for $d=20$
[see Fig.~\ref{fig:cn-offsite}(c)], both the $P(\mu )$ curve and the
corresponding shape transition of the off-site GSs are close to their
counterparts in the local model. A smooth transition of the GS solutions
from the single-peaked shape to a flat-top one, and then to the double-peaked
(in-phase) shape can be traced, in the latter case.

However, for a smaller degree of the nonlocality, which makes the
self-defocusing nonlocal response comparable to the Kerr nonlinearity, the
analysis reveals the existence of more than one branches of the off-site
GSs. For small values of $d$, such as $d=0.05$ shown in Fig.~\ref%
{fig:cn-offsite}(c), one branch (the bold red curve in the figure) extends
continuously from the upper to lower edge of the bandgap. The modal profile
for this branch remains double-humped, such as the one shown by dashed lines
of Fig.~\ref{fig:cn-offsite}(a) and (b) for markers B and D in (c),
respectively. Besides this branch of the double-humped GS modes, there is a
separate branch representing solutions with a lower formation power and
single-humped profile, as shown by solid lines in Fig.~\ref{fig:cn-offsite}%
(a) and (b) for markers A and C in panel (c).

\begin{figure}[tbp]
\centering
\includegraphics[width=8.3cm]{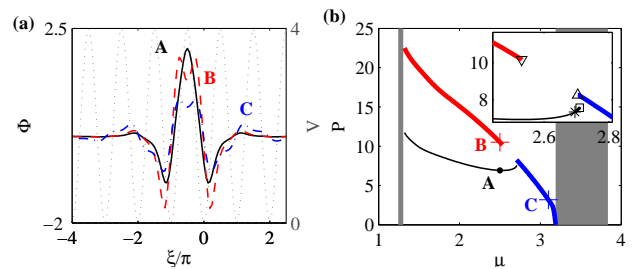}
\caption{(Color online) (a) Off-site GSs under the competing local
self-focusing and nonlocal self-defocusing nonlinearities, corresponding to
the points A ,B and C in panel (b), where the $P(\protect\mu )$ curves for
the nonlocality degree $d=2$ are shown. The thin and thick lines distinguish
portions corresponding to single- and double-peaked modes, respectively.
Inset in (b) is a blow-up of the region around the marked points.}
\label{fig:cn-offsite-2}
\end{figure}

Thus, the single- and double-peaked GSs, which constitute parts of the single
GS family in the local model, split into two disjoint families in the model
with the appreciable competition between the local and nonlocal
nonlinearities. In the intermediate case, corresponding to a moderate degree
of the nonlocality, such as for $d=2$ shown in Fig.~\ref{fig:cn-offsite-2}%
(b), the $P(\mu )$ curve for the double-humped GSs break into two segments:
one starts from the lower edge of the bandgap and ends at a cutoff point
corresponding to marker B in the figure, and the other starts from the upper
edge and ends at another cutoff point, which is designated by marker $%
\triangle $. The cutoff points may be accounted for by bifurcations
involving additional higher-order modes, which we did not aim to
find in this work dealing with fundamental on- and off-site GSs.
Another $P(\mu )$ branch chiefly represents the single-humped GSs,
but it also contains a portion to the right of point A, which
corresponds to double-humped modes.

The inset of Fig.~\ref{fig:cn-offsite-2}(b) clearly shows that there are
three branches of double-humped modes. The first one bifurcates from the
upper edge of the bandgap and ends at the point marked $\triangle $; the
second one extends from the lower edge of the bandgap and terminates at
point labeled by $\triangledown $; the final branch bifurcates as the
flat-top solution from the point marked the asterisk ($\ast$) and terminates at the site marked $\square $. %
The first two branches that bifurcate from either edge of the bandgap abruptly terminate inside the bandgap, where the characteristic width of the nonlinear response is larger than or comparable to half the lattice period, which makes the balance between the nonlinear and lattice-induced effects impossible. The last branch, which starts as the flat-top mode, ends due to the divergence of the total power as a result of vanishing nonlinearity, similar to cutoff~considered in Ref. \cite{eig-cutoff}. %

\begin{figure}[tbp]
\centering
\includegraphics[width=8.3cm]{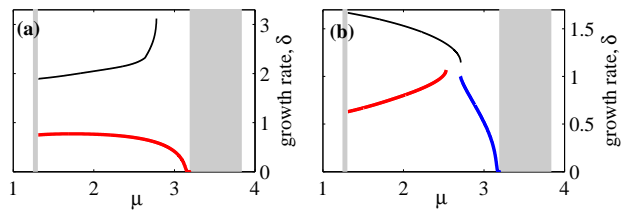}
\caption{(Color online) Eigenvalues of small-perturbation mode upon the Off-site GSs under the competing local self-focusing and nonlocal self-defocusing nonlinearities, of which the degree of non-localities are (a) $d=0.05$ and (b) $d=2$. The color and line style to distinguish each branch is as is defined in the corresponding $P(\protect\mu )$ curves in in Fig.~\ref{fig:cn-offsite}(a) and Fig.~\ref{fig:cn-offsite-2}(b), respectively. }
\label{fig:cn-offsite-stab}
\end{figure}

Even though off-site GSs are in general unstable both in local~\cite{bifur-gap-2004} and nonlocal~\cite{joa-yylin} nonlinearities despite the inverted "anti-Vakhitov-Kolokolov" criterion, the corresponding instability growth rate is proportional to the GS's power after a certain threshold value~\cite{osc-ins-1994,joa-yylin}. %
Due to the unstable nature of off-site GSs, 
we study the instability of GSs by the linear stability analysis for the branches off-site GSs and identify the final state of these off-site GSs by beam propagation simulation. 
The linear stability spectrum in Fig.~\ref{fig:cn-offsite-stab} shows the amplitude growth rate of the small-perturbation eigen-modes found upon those off-site GS's revealed previously. %
For a smaller degree of non-locality, $d=0.05$, illustrated in Fig.~\ref{fig:cn-offsite-stab}(a) the two distinct branches are unstable and the corresponding eigen-modes upon single-peaked GSs have higher amplitude growth rate which diverges near its cutoff point in the band gap region. %
The GSs on these two branches  collapse fast as they propagates due to a larger  growth rate. %
Likewise, in Fig.~\ref{fig:cn-offsite-stab}(b) when a moderate degree of non-locality, $d=2$, is considered, the single-peaked GSs branch acquires a stronger instability than that of the two separated double-peaked GSs branches. %
The two double-peaked GSs branches, though both feature the worst instability near their cutoff points in the band gap region, reflect very different relations to the corresponding GS's power, which we believe is stemmed from the cutoff of the branches.

\begin{figure}[h]
\includegraphics[width=9.3cm]{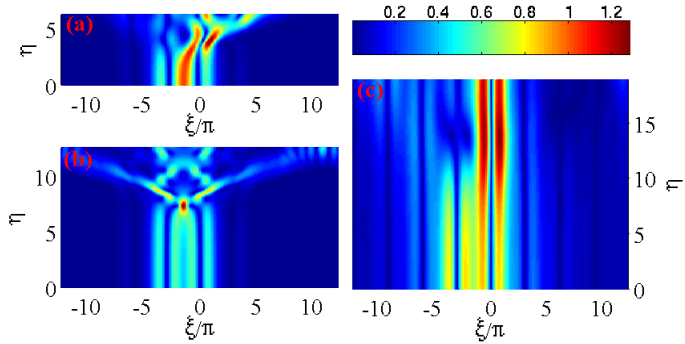}
\caption{(Color online) \label{fig:cn-offsite-porop} Three examples of the beam propagation simulations for off-site GS solutions shown in Fig.~\ref{fig:cn-offsite-2}(a), in which modes A, B and C correspond to intensity plots in (a), (b), and (c), respectively.}
\end{figure}

The beam propagation simulations further illustrate that GSs of the single-peaked branch and double-peaked modes with a higher energy  fall into collapsed states. %
Nevertheless, mode conversions from unstable off-site GSs into stable on-site GSs are observed for the branch plotted in blue in Fig.~\ref{fig:cn-offsite-2}(b). %
Three examples of beam propagation simulations are illustrated, resulting in either collapsed states Fig.~\ref{fig:cn-offsite-porop} (a) and (b) or a mode conversion behavior Fig.~\ref{fig:cn-offsite-porop} (c). %
Information such as mode transition or conversion is beyond what linear stability analysis may reveal. %
Even though it is also believed that interesting dynamical behavior associated with the GSs can be delineated through a direct beam propagation simulation, yet to be more focused, a thorough investigation of propagation behavior goes beyond the scope of this work. %

The nonlocal nonlinearity competing with the local Kerr term not only
reduces the strength of the nonlinearity but also tailors the effective
nonlinear response to induce binding forces outside of the lattice-potential
barriers, which is a more favorable environment for the existence of
double-peaked modes. Therefore, the three GS branches outlined above are
formed owing to the interplay of the potential barrier created by the
lattice potential and the binding potential induced by the effectively \emph{%
reduced} Kerr nonlinearity. In this case, the GS solutions belonging to the
branch originating from the lower edge of the bandgap have a larger
amplitude and are tighter localized. The width of the corresponding response
range of the self-defocusing nonlocal nonlinearity is larger than that of
the Kerr response. Such a double-peaked branch cannot exist in the
strong-nonlocality limit, and we numerically find that values of the
nonlinearity degree supporting this branch are bounded by $d<9.2$. As the
power decreases (the propagation constant increases), the widening of the
the GS mode makes the overall nonlinear response effectively local,
suppressing the capability of the nonlocal nonlinearity to tailor its
response to a shape necessary for supporting the solitons. Then, when the GS
width becomes smaller than or comparable to half the lattice period, the
double-peaked modes cease to exist because the balance between the lattice
potential and nonlinearity-induced perturbation of the refractive index
supports only single-peaked modes.

\begin{figure}[tbp]
\centering
\includegraphics[width=8.3cm]{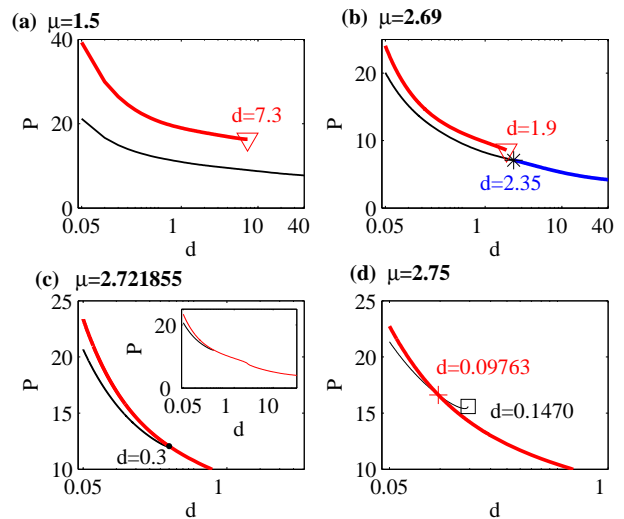}
\caption{(Color online) The power, $P$, versus the degree of the
nonlocality, $d$, at fixed values of the propagation constant: $\protect\mu %
=1.5$ (a), $2.69$ (b), $2.721855$ (c), and $2.75$ (d), respectively. The
inset in (c) is the blow-up of the region in the vicinity of the merger point.
}
\label{fig:P-d}
\end{figure}

To present a clear description of shape transitions for the off-site
GSs, we replot the relationship of power $P$ versus the nonlocality
degree, $d$, for a fixed propagation constant $\mu $ in
Fig.~\ref{fig:P-d}. For a smaller propagation constant, such as $\mu
=1.5$ in Fig.~\ref{fig:P-d}(a), the power necessary to support a
single-humped off-site GS is always lower than that of its
double-humped counterpart, for all values of $d$. Moreover, above a
critical degree of the nonlocality, $d=7.3$ in this case, only a
single-humped GS can be found. Moving $\mu $ into the center of the
bandgap -- for instance, taking $\mu =2.69$ in Fig.~\ref{fig:P-d}(b)
-- the critical degree of the nonlocality reduces to $d=1.9$, and,
above another critical value, $d=2.35$, the single-humped mode
transforms into a double-humped one,
as in the local model. Increasing the value of the propagation constant to $%
\mu =2.721855$, the two $P(d)$ curves merge at $d=0.3$ in in Fig.~\ref%
{fig:P-d}(c). For a larger value of the propagation constant, such as $\mu
=2.75$ in Fig.~\ref{fig:P-d}(d), the two curves intersect at a critical
value $d=0.09763$. Above this critical point, the power for the
double-humped off-site GS becomes lower than for a single-humped one. In
this case, the profile of the off-site GS mode can be switched from
single-humped into double-humped by adjusting the nonlocality degree, $d$.

\begin{figure}[tbp]
\centering
\includegraphics[width=8cm]{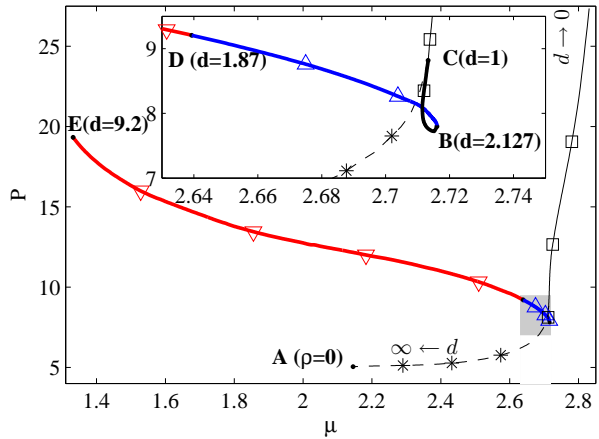}
\caption{(Color online) The diagram in the plane of the propagation constant
($\protect\mu $) and power ($P$), for the gap solitons, as different values
of the nonlocality degree, $d$. The inset is a blowup of the gray region.
Route AC corresponds to the transition point marked by $\ast $ in Fig.~%
\protect\ref{fig:cn-offsite}; routes BC and BDE correspond to the
double-peaked modes marked by $\bigtriangleup $ and $\bigtriangledown $ in
Fig.~\protect\ref{fig:cn-offsite-2}, respectively; the route above point C
corresponds to the single-peaked solution. The marked points are A with $%
\protect\rho =0$, and B, C, D and E with $d=2.127$, $1$, $1.86911$, and $9.2$%
, severally.}
\label{fig:phase-diagram}
\end{figure}

To facilitate the understanding of the present picture, we consider the
plane of the propagation constant $\mu $ and power $P$ for the GS solutions
in~Fig.~\ref{fig:phase-diagram}, varying the nonlocality degree $d$. We
start by tracing the evolution of the point of the transition from
single-peaked to the double-peaked shape, marker by the asterisk ($\ast $) in
Fig.~\ref{fig:cn-offsite-2}(c). In the absence of the competing nonlocal
nonlinearity, i.e., at $\rho =0$ in Eq. (\ref{eq:gp1}), the transition point
is $(\mu =2.145,P=5.0460)$, labeled by A in Fig.~\ref{fig:phase-diagram},
which also corresponds to the limit of $d\rightarrow \infty $. As the
nonlocality degree drops to a critical value, $d=1$ at point C $(\mu
=2.7133,P=8.8184)$, the transition point ceases to exist (i.e., only sharply
peaked single-humped modes are supported by the system), merging into to the
end point of the doubled-humped-mode branch, marked by $\square $ in Fig.~%
\ref{fig:cn-offsite-2}. Increasing the nonlocality degree from $d=1$ at
point C, the end point $\square $ in Fig.~\ref{fig:cn-offsite-2} merges into
the other end point $\bigtriangleup $ in Fig.~\ref{fig:cn-offsite-2} at $%
d=2.127$. The latter merger happens at point B in~Fig.~\ref%
{fig:phase-diagram}.
This is the end point of the family of the double-peaked modes [an example
corresponds to the point marked by $\triangle $ in the inset of Fig.~\ref{fig:cn-offsite-2}%
(b)] which originates from point B in Fig.~\ref{fig:phase-diagram} at the
critical nonlocality $d=2.127$, and extends toward point D, which
corresponds to $d=1.86911$, where it merges into a new branch of the
double-peaked modes emerging (as long as $d<9.2$) from the lower edge of the
bandgap at point $(\mu =1.3047,P=19.3262)$. %

\section{Conclusion}

\label{sec6} In this work, we aimed to study GS (gap-soliton) solutions in
the first finite bandgap of the periodic potential, with the nonlinearity
represented by the competing local self-focusing and nonlocal
self-defocusing terms. The two terms are balanced so that, in the limit of
the zero nonlocality radius, they exactly cancel each other. While keeping
the effective interaction self-attractive, the existence, stability, and
bifurcation for on-site and off-site modes were analyzed numerically. Due to
the opposite signs of the local and nonlocal nonlinearities, an increased
power was required for the formation of both the on-site and off-site GSs.
The competing nonlinearities induce a region where stable on-site modes
obeying the "anti-Vakhitov-Kolokolov criterion" near the
upper edge of the bandgap region become unstable. For unstable off-site GS
modes, which remain unstable under competing local and nonlocal nonlinearities,
a complex bifurcation pattern with cutoff points was found and
explained in terms of the transitions between the single-humped, flat-top,
and double-humped shapes. By tracing the evolution with the change of the
nonlocality degree, we have shown that it is possible to switch different
off-site GS modes by manipulating the nonlocal interaction against the local
Kerr nonlinearity.

\section*{Acknowledgement}

This work was partly supported by the National Science Council of Taiwan
with contrasts NSC 95-2112-M-007-058-MY3, NSC 95-2120-M-001-006 and NSC
98-2112-M-007-012.

\end{document}